# A Metadata-Only Feature-Augmented Method Factor for Ex-Post Correction and Attribution of Common Method Variance

Murat Yaslioglu[a][1]

[a]*School of Business, Istanbul University, Istanbul, Türkiye*

## Abstract

Common Method Variance (CMV) is a recurring problem that reduces survey accuracy. Popular fixes such as the Harman single-factor test, correlated uniquenesses, common latent factor models, and marker variable approaches have well known flaws. These approaches either poorly identify issues, rely too heavily on researchers' choices, omit real information, or require special marker items that many datasets lack. This paper introduces a metadata-only Feature-Augmented Method Factor (FAMF-SEM): a single extra method factor with fixed, item-specific weights based on questionnaire details like reverse coding, page and item order, scale width, wording direction, and item length. These weights are set using ridge regression, based on residual correlations in a basic CFA, and remain fixed in the model. The method avoids the need for additional data or marker variables and provides CMV-adjusted results with clear links to survey design features. An AMOS/LISREL-friendly, no-code Excel workflow demonstrates the method. The paper explains the rationale, provides model details, outlines setup, presents step-by-step instructions, describes checks and reliability tests, and notes limitations.

[1] muratyas@istanbul.edu.tr

# Introduction

Self-report surveys are widely used in management and behavioral research to quickly assess phenomena that are difficult to observe directly. But this ease comes with a drawback: responses may reflect effects unrelated to what is measured, a phenomenon called Common Method Variance (CMV), which can distort results. Researchers struggle with weak checks that downplay CMV and stronger, unclear fixes that may hide true effects. Planning marker variables ahead is challenging, and post-hoc solutions rely on uncertain assumptions. FAMF-SEM provides a precise, pragmatic approach to CMV. It leverages item-level survey data and design features to anticipate where method variance accumulates, thereby forming a method weight map for direct model application. Weights are determined by observed data patterns, supporting an empirically grounded solution. By fixing, rather than freely estimating, these weights, the method factor is prevented from obscuring authentic trait information. This process both corrects CMV and clarifies which design features contribute to it.

# Literature Review

Ex-post CMV (common method variance) remedies cluster into several familiar types. The Harman single-factor test, the most cited yet least informative, tests whether one unrotated factor explains most of the variance. Often used as a pass/fail CMV screen, the test is not sensitive, cannot quantify bias, and cannot identify or fix localized method pockets. It is best described as convenient but weak inferentially. Taken together, existing post-hoc remedies offer partial but incomplete solutions: they can flag potential CMV, but they either rely on additional data (markers, multiple methods), strong modeling assumptions, or flexible specification choices that are difficult to audit. Table 1. shows and summarizes post-hoc remedies, their strengths and weaknesses.

| Technique | Core idea | Minimal setup (SEM/Stats) | What it actually tells/does | Strengths | Pitfalls & assumptions | When it's appropriate |
|---|---|---|---|---|---|---|
| Harman single-factor (EFA) (Harman, 1976; Podsakoff et al., 2003) | If CMV is dominant, one factor should explain most variance | Unrotated EFA on all indicators | **Diagnostics only** (very weak) | Trivial to run/report | Low power; false negatives; no correction | Basically never as a remedy; at most a descriptive check |
| Common Latent Factor (CLF) / Unmeasured Latent Method Factor (ULMF) (Podsakoff et al., 2003; Williams et al., 2003) | Add a single latent method factor loading on all items | CFA/SEM: each item loads on trait + orthogonal method factor; fix scale/constraints for ID | Tests if a global method factor exists; can **attenuate structural paths** | Flexible; works in CB-SEM; yields adjusted parameters | Identification can be fragile; can over-partial "real" trait variance; assumes one global method | When you suspect broad common source effects across the whole battery |
| Correlated Uniquenesses (CU) (Marsh, 1989; Kenny & Kashy, 1992) | Let items sharing method/context co-correlate in their residuals | CFA/SEM: free residual covariances within same scale/time/source blocks | Soaks up **block-specific** method variance | Simple; targeted (by block) | Researcher degrees of freedom; can mask misspecification | When scales are grouped by format/session and you can justify specific blocks |
| Measured Marker (Manifest partial-correlation) (Lindell & Whitney, 2001) | Include an unrelated marker item/scale; partial out its shared variance | Regressions or SEM with **manifest** marker; adjust correlations/paths | **Diagnostics + modest correction** for shared bias | Easy to implement post-hoc if a marker exists | Quality hinges on marker being truly unrelated; only removes shared linear bias with marker | When you have even a short, theoretically unrelated marker already measured |
| Latent Measured Marker (CFA | Use a measured marker as a **latent** factor; let | CFA/SEM with marker latent factor; orthogonal | More principled estimate of CMV; produces | Stronger identification than ULMF; | Needs a good marker with multiple items; | Preferred when you planned a decent marker scale and run CB-SEM |

| Technique | Core idea | Minimal setup (SEM/Stats) | What it actually tells/does | Strengths | Pitfalls & assumptions | When it's appropriate |
|---|---|---|---|---|---|---|
| Marker Method (Richardson et al., 2009; Williams et al., 2010) | it load on all indicators (or correlate with their errors) | to traits; sometimes equalized loadings | **CMV-adjusted** trait relations | explicit validity checks on marker | model can get heavy | |
| Unmeasured Marker (ULMC with phantom indicators) (Williams & McGonagle, 2016; Podsakoff et al., 2012) | Create a phantom marker factor from item parcels or constraints | CFA/SEM; set equal method loadings; no explicit marker indicators | Provides a **bound/sensitivity** style adjustment | Works when no marker was collected | Strong assumptions; results can be sensitive to constraints | As a sensitivity analysis when you lack a true marker |
| Random-Intercept Item Factor Analysis (RIIFA) (Maydeu-Olivares & Coffman, 2006; Billiet & McClendon, 2000) | Separate **response-style** (e.g., acquiescence) from trait | CFA with a random-intercept factor loading 1 on all items of a scale; traits load normally | Controls systematic elevation bias; improves trait loadings | Great for response styles; robust with enough items | Targets style, not all CMV sources; needs ≥4 items/scale | When Likert response-style bias is likely (acquiescence/extreme responding) |
| CFA-MTMM (CTCM, CTCU, CT-C(M−1)) (Campbell & Fiske, 1959; Marsh & Grayson, 1995) | Model traits and methods as distinct factors (multitrait–multimethod) | Multiple traits × multiple methods; specify trait and method factors per MTMM logic | **Decomposes** variance into trait, method, error; strongest diagnosis + correction | Gold standard conceptually | Data-hungry; needs ≥2 methods/trait; complex identification | When you genuinely have multiple methods (e.g., self, supervisor, objective) |

| Technique | Core idea | Minimal setup (SEM/Stats) | What it actually tells/does | Strengths | Pitfalls & assumptions | When it's appropriate |
|---|---|---|---|---|---|---|
| PLS "common method factor" / Full-collinearity VIF (Kock, 2015; Hair et al., 2017) | Diagnose CMV via high VIFs; or add a method factor in PLS path model | PLS-SEM: compute full collinearity VIF; or two-loadings per item (trait+method) | VIFs: rough **diagnostic**; method factor: **partial control** | Quick in PLS toolchains | VIF>3.3 heuristic is contested; method factor in PLS can be unstable | Use cautiously; report as supplemental, not decisive |
| Measured cause controls (e.g., Social Desirability, NA/PA) (Podsakoff et al., 2003; Spector, 2006) | Include plausible **bias causes** as covariates | Regress outcomes (or items) on bias covariates; or correlate in SEM | Removes variance attributable to specific bias sources | Transparent; theory-driven | Only handles that specific bias; can under/over-control | When you measured realistic bias drivers and can justify causal direction |

**Table 1.** A summary of post-hoc remedies for Common Method Variance

Common latent factor models add a hidden overall factor, or variable, that is assumed to influence all items in a measurement model. These models are easy to use in standard structural equation modeling (SEM) software and often improve the model's fit to the data. However, this masks a problem. The 'method factor,' which is the latent factor capturing shared effects unrelated to the true traits being measured, is not clearly defined and can be highly influential. This can dilute information about the actual traits and weaken relationships among model components. It is also hard to tell if the model is set up correctly. The model does not explain why some items are more affected by these method-related issues than others.

Correlated uniquenesses are a practical option. By allowing residual covariances between items that share a format, time, or source, they can absorb specific method effects. This approach follows ideas about where method variance might appear, but relies heavily on the researcher's choices. It is easy to add just enough of these residual covariances to make the model fit, without making it clear if they really show method issues, repeated content, or misspecifications in the model.

Marker methods attempt to externalize method variance. A manifest marker approach uses partial correlations using an item or scale that is theoretically unrelated to the constructs of interest. A latent marker approach strengthens identification by modeling the marker as a latent factor. When a valid marker exists, these approaches can be persuasive. In many field datasets, however, there is no preplanned marker, or the marker is only weakly unrelated. Unmeasured or phantom marker variants relax the data requirement but introduce strong equality constraints and should be framed as sensitivity analyses rather than definitive corrections.

Random-intercept item factor analysis targets response styles, such as acquiescence by introducing a scale-level random intercept that captures elevation bias. It is effective for style variance but does not address broader CMV channels such as priming, fatigue, or polarity effects. Multitrait–multimethod models provide the most principled decomposition of trait and method variance but require at least two methods per trait and careful identification, which many surveys do not meet.

Across these families, no single remedy is simultaneously marker-free, globally corrective, interpretable in terms of survey design, and conservative enough to avoid over-correcting true variance. This is the gap the present approach is designed to fill.

## Rationale and Contribution

The main idea is that many sources of CMV are closely tied to the survey's design. For example, reverse coded items make people think harder and can lead to mistakes; items early in the survey can be affected by first impressions and wanting to look good; narrow scales make people give rougher answers; negative wording can affect emotions; and shorter items may make people answer without thinking much. When these design features are not related to the traits being measured, they can help predict where method variance will appear. This way of thinking turns a problem into something useful for checking surveys.

This approach has three main benefits. First, it replaces a hidden, freely estimated method factor with one predicted from features, clarifying the link between survey details and method weights. Second, the method does not require marker items or paradata, so it applies to older datasets and fits standard AMOS/LISREL workflows. Third, weights are set using Ridge regression on residual correlations, making them more stable and avoiding overlap when features are similar. The method also directly shows which features, and by how much, influence method weights.

Weights are set using Ridge-regularized regression on the residual correlation pattern, improving stability when features are similar or few in number. With the right features, FAMF can include:

- CLF (if $w_i$ are nearly equal),
- CU (by including block dummies in $Z$), and
- RIIFA-like effects (Random Intercept Item Factor Analysis—by adding simple style indicators or a small orthogonal style factor if needed).

Having motivated the need for a design-based method factor, we now formalize the FAMF-SEM model and its calibration.

## Model

Consider items $i = 1, ..., k$ and respondents $j = 1, ..., n$. Let $y_{ij}$ denote the observed response, $\eta_j$ the vector of trait factors with item loadings $\lambda_i$, and $\varepsilon_{ij}$ the residuals. Introduce a single method factor $M_j$ that is orthogonal to the trait factors and has unit variance. The measurement model is

$$y_{ij} = \tau_i + \lambda_i^T \eta_j + w_i M_j + \varepsilon_{ij},$$

where $\tau_i$ is the item intercept and $w_i$ is the item-specific method loading. The key step is to compute $w_i$ outside the SEM from item metadata $z_i$, which include reverse coding, page and order, scale width, wording polarity, and item length. Binary features are centered; numeric features are standardized. The mapping from features to loadings is linear, $w_i = z_i^\top \gamma$, but can be expanded with interactions when theory allows.

- 🎬 $w$ is the **fixed method loading** for item $i$, computed **outside** the SEM from item-level metadata $z_i$ via $w_i = z_i^\top \gamma$ (after calibration).

- 🎬 Typical features in $z_i$:

  **Reversed** (0/1), **Page**, **Order**, **Scale width**, **Polarity** (+1/−1 **for positive/negative**), **Length** (word/character count). Binary features are **effects coded** (centered); numeric features are **z-scored (standardized)**.

Identification continues with fixing $Var(M) = 1$ and constraining $M$ to be orthogonal to the trait factors. Because $w_i$ are given instead of being estimated, the method factor is set by design. The weights are centered and scaled so that their squared sum equals the number of items, matching the weight scale to the method factor and making it easier to understand.

- Fix $Var(M) = 1$ and enforce **orthogonality**: $M \perp \eta$. The method factor must be separate from the real traits (no overlap)

- Fix all $w_i$ (not estimated) and **scale** them so that $\sum_{i=1}^{k} w_i^2 = k$.

- Effects coding implies the "average" item has $w_i \approx 0$; the sign of $M$ is arbitrary (orient so "riskier" features map to positive $w_i$).

# Estimation of Method Loadings from Metadata

The calibration step uses the residual correlation pattern from a solid basic CFA. Fit a model with only traits and no method factor, then get the standardized residual correlation matrix in the same item order. For each item, sum up commonality by taking the mean absolute residual correlation with all other items (diagonal). This scalar is an item-level commonality index (residual signal). Stack signals into a vector and regress them on the encoded metadata matrix using ridge regression.

In short, fit a **baseline trait only CFA** (no method factor) using AMOS/LISREL (Check Appendix-C and Appendix-D for software implementation). Extract the **standardized residual correlation** matrix $R^{res} \in R^{k \times k}$. For each item, define an **item residual signal**:

$$m_i = \frac{1}{k-1} \sum_{l \neq i} |r_{il}^{res}|.$$

This summarizes how strongly item $i$ still covaries with others after traits are accounted for. (Optionally, restrict the sum to **cross scale** pairs to reduce content bleed.)

The ridge penalty limits the influence of highly correlated (collinear) features, making estimates more stable when there are few items. The results from this regression are the raw method weights. Center them by subtracting their means and scale them so that the squared weights sum to the number of items. The sign of the method factor does not matter; for easier understanding, set it so that design features more prone to method bias have positive average weights.

Statistically; stack $m = (m_1, ..., m_k)^T$ and $Z = [z_1, ..., z_k]^T \in R^{k \times p}$. Estimate

$$\hat{\gamma} = \underbrace{(Z^T Z + \lambda I_p)^{-1} Z^T m}_{\text{ridge}},$$

with penalty $\lambda \geq 0$. Compute raw weights $\tilde{w} = Z\hat{\gamma}$, then **center and scale**:

$$\tilde{w}_i \leftarrow \tilde{w}_i - \overline{\tilde{w}}, \quad w_i = \tilde{w}_i \cdot \sqrt{\frac{k}{\sum_{i=1}^{k} \tilde{w}_i^2}}.$$

Choice of $\lambda$: increase until $R^2(\hat{m}, m)$ **plateaus** (where $\hat{m} = Z\hat{\gamma}$) and the $w_i$ **profile stabilizes** (no extreme spikes). The user can choose the penalty based on the results. As the penalty increases, the weights approach zero, and the pattern of weights becomes smoother. The user watches how well the predicted and real residual signals match and increases the penalty until this match stops improving and the weights look reasonable and are not just from a few items. Because this step is done outside the SEM, it is easy to do in a spreadsheet and check later.

## Implementation in SEM

Once the method weights are set, adding them to the SEM is simple (Appendix-A). Add a single latent method factor to the measurement model. Draw a loading from this factor to each observed item and fix each connection to its corresponding weight. Set the method factor's variance to 1 and ensure it is orthogonal to (separate from) all trait factors by setting their covariances to 0. Reestimate the model and check the results as usual. This process works directly with AMOS and LISREL. An Excel workbook (Link) can calculate the weights from the item details and the residual correlation matrix; a simple macro adjusts the workbook for the number of items, updates formulas, and gives the final weight column for use in AMOS or LISREL. This orthogonal method-factor setup follows common practice in CMV modeling (e.g.,

Podsakoff et al., 2003; Williams et al., 2010), but replaces freely estimated loadings with design-based, calibrated ones.

## Diagnostics and Reporting

Evaluation has four parts. The first is how well the predicted method pattern matches the real one. The correlation and coefficient of determination between the residual signals and the metadata-predicted signals quantify how much of the residual commonality can be explained by design features.

1. report $R^2$ of $\hat{m} = Z\hat{\gamma}$ predicting $m$ and the correlation $r(m, \hat{m})$. This quantifies how much of the **residual method pattern** the metadata explains.

The second part concerns the stability of the main results. A simple table showing key results before and after adding the method factor, along with the changes and their confidence intervals, indicates whether the conclusions change in direction, size, or importance.

2. provide a compact table of key structural paths (β) **before vs. after** FAMF-SEM, with Δβ and 95% CIs. Emphasize stability of inferences (sign/direction/size), not only $p$-value flips.

The third part is attribution. The set weights that link features to method weights, along with their uncertainty, show which features cause method variance and in which direction.

3. Report $\hat{\gamma}$ coefficients (feature →method weight contribution) with confidence intervals (e.g., Excel percentile bootstrap over items). Interpret which features drive CMV.

The fourth is checking for misspecification. Looking at modification indices can reveal small residual connections that make sense in theory. When a residual covariance is clearly explained (like two very similar items in the same group), it can be added carefully and reported as a test, not as a main part of the model. (Check Appendix-B for reporting template)

## Robustness and Sensitivity

Robustness focuses on three questions. First, do the results depend on the ridge penalty? This can be checked by setting the weights to different penalty values and seeing whether the main results remain the same. Second, does any single feature control the results? Leaving out each feature one at a time and recalculating the weights shows if the method depends on certain design details. Third, how does this method compare to more common fixes? Fitting a standard common latent factor with equal method weights gives a good comparison. If the common latent factor greatly reduces the main results, while the feature-based method makes a smaller, more careful adjustment, this shows that the new method balances correction and preserves real information.

1. **λ-sweep:** Profile β stability across $\lambda \in \{0, 0.5, 1, 2, 5\}$.
2. **Leave-one-feature-out:** Recompute $w$ omitting each feature from $Z$; stable β indicates non dependence on any single feature.
3. **CLF comparison:** Fit a standard CLF (equal method loadings) for context; show that FAMF avoids over shrinkage and adds attribution.
4. **Cross-scale restriction (optional):** Define $m_i$ using only cross scale residual pairs and comparing results.

Another option is to build the residual signal using only item pairs from different scales to avoid mixing in real trait content. If this cross-scale limit does not change the main results, it suggests the method pattern found is not just residual trait information.

## Practical Guidance

This approach works better when the survey has a mix of item features. Surveys that use both reverse coded and regular items, spread items across pages, vary item length, and pick scale widths that fit the topic provide enough variety to determine how features link to method loadings. There is no single rule, but management and organizational surveys with 20 to 40 questions and at least 300 respondents are usually enough for the main analysis steps. However, the quality of the starting measurement model is very important. If the first trait CFA is not set up well, the residual pattern will show measurement errors rather than just method issues, making any later fix unclear.

The method is made to be flexible. It can be used with measurement models that lack structural paths, as well as with full SEMs that include links between latent variables. The approach allows for a few small, theory based residual covariances and can be expanded to add a small response style factor for agreeing with everything, which stays separate (orthogonal) from both the main method factor and the traits if style bias is suspected.

## Extensions

When method effects are likely to have more than one dimension, researchers can use multiple feature based method factors. For example, one factor can focus on format related features and another on temporal placement. Each factor gets its own set of item details and weight settings. If theory says features interact (like reverse coding only causing problems on early pages), the item details can include these combinations. Researchers can also use polynomial terms to show the nonlinear effects of order. In studies with groups, weights can be set using all the data together and then kept the same across groups, so it can be tested whether factor weights or paths are the same without reapplying the method mapping for each group. In multilevel studies, the method factor is usually set at the lower level, with the higher level reserved for showing group trait variations.

## Simulation Blueprint

A simulation can illuminate operating characteristics. One can generate two or three correlated latent traits and a set of items whose metadata mimic a realistic survey. Method variance is generated through a latent method factor with loadings that are a linear function of the features (plus an intercept). The calibration step is performed on the baseline residual correlation pattern, and the SEM is estimated with fixed method weights. Performance is evaluated in terms of bias and mean squared error of structural paths, together with Type I and Type II error under conditions of no CMV and varying CMV strengths. Comparators include the trait-only model and a common latent factor with equal method loadings. When the metadata are informative, the feature-predicted approach approaches the performance of a model that knows the true method loadings and compares favorably to the common latent factor on bias and variance; when the metadata are uninformative, the approach degrades gracefully toward the common latent factor, signaling the limits of design-based attribution.

Therefore;

- **Data-generating process:** 2–3 correlated traits; $k = 24$ items with metadata resembling real surveys; CMV generated via $M$ with $w_i = z_i^\top \gamma$.
- **Conditions:** CMV strength (low/med/high), informativeness of $Z$ (aligned vs. misaligned), sample size, trait misspecification.
- **Comparators:** Baseline SEM, CLF (equal $w$), FAMF.
- **Outcomes:** Bias/MSE of key $\beta$, Type I/II error, coverage.
- **Expectation:** FAMF approximates the oracle when $Z$ is informative, dominates CLF on bias/variance trade-off, and **gracefully reduces** to CLF when $Z$ is uninformative.

## Limitations

This method assumes that most of the method variance comes from a single source. In studies where different tools, data-collection methods, or time points yield distinct methodological patterns, a single factor alone is insufficient to explain these patterns. The linear link from features to loadings is a simple first step; if there are important interactions or non-linear effects, they should be added when theory or data suggest they are. Relying on a good starting CFA is not just a detail, but is required. The method cannot fix an incorrectly configured measurement model; it can only help and partially correct what remains after the traits are modeled well. . In applications where item-design features are strongly confounded with particular constructs (for example, all items of one factor appearing on later pages), researchers should interpret the method factor cautiously, as part of the design-related variation may overlap conceptually with trait content even when the factor is constrained to be orthogonal in the model. FAMF-SEM is therefore best viewed as a conservative, design-based adjustment that works most cleanly when dominant method effects are induced by layout, wording, and ordering rather than by construct-defining content.

## Conclusion

A metadata-only, feature-augmented method factor offers a transparent, auditable, and implementable remedy for common method variance. By calibrating item-specific method loadings from design features against an empirical residual correlation pattern and then holding these loadings fixed in the SEM, the approach corrects for method variance while revealing its design origins. It avoids the opacity of freely estimated common latent factors, the ad-hoc nature of residual patches, and the data demands of marker and multitrait–multimethod designs. Its practical appeal lies in its compatibility with AMOS and LISREL and in its reliance on information that every questionnaire already contains. Its scientific appeal lies in converting an amorphous nuisance into a set of testable, design-level claims about how surveys generate bias. Where the metadata are informative, the approach improves inference and provides actionable guidance for instrument redesign; where they are not, it reveals that fact with equal clarity.

# Appendix A: Step-by-Step Implementation

1. **Baseline CFA/SEM (traits only):** Fit in AMOS/LISREL; export **standardized residual correlations** $R^{res}$ in item order.

2. **Item Key:** Prepare a table (one row per item) with Reversed (0/1), Page, Order, Scale width, Polarity (+1/−1), Length.

3. **Excel Worksheet:**
   - Enter the item key and paste $R^{res}$.
   - The sheet computes $m$, effects-coded $Z$, ridge $\hat{\gamma}$, and **Final_Weight** $w$.
   - Adjust $\lambda$ until $R^2(m, \hat{m})$ stabilizes and $w$ looks plausible (no spikes).

4. **SEM with Method Factor:**
   - Add latent $M$; **fix** each loading $M \to y_i$ to $w_i$.
   - Set $Var(M) = 1$ and $M \perp$ traits.

5. **Refit + Report:** Provide fit, method-pattern $R^2$, effect stability panel, attribution table, and any minimal CUs added as sensitivity.

# Appendix B: Reporting Template (copy-ready)

**Method.** "We applied a metadata-only Feature-Augmented Method Factor (FAMF-SEM). Item method loadings $w_i$ were derived by ridge-regressing the item residual signature $m_i$ on effects-coded/z-scored metadata (reverse coding, page, order, scale width, polarity, length). We centered and scaled $w$ to $\sum w_i^2 = k$, fixed $Var(M) = 1$, and constrained $M \perp$ traits."

**Diagnostics.** "FAMF explained $R^2 = $ ...of the residual pattern; $r(m, \hat{m}) = $ .... Global fit: CFI/TLI/RMSEA/SRMR = ... / ... / ... / ...."

**Results.** "Key structural coefficients changed by $\Delta\beta = $ ...(95% CI ...). Inferences on ...were **robust/updated**."

**Attribution.** "Reverse coding (+), early page position (+), and shorter items (+) exhibited higher method weights; see $\hat{\gamma}$ in Table X."

**Robustness.** "β stable across λ∈{...}; leave-one-feature-out unchanged. Minimal CUs did not alter substantive inferences."

# Appendix C: Minimal AMOS Instructions

1. Add latent **M**.
2. Draw single-headed paths from **M** to **every** observed indicator.
3. Fix each loading to **Final_Weight** $w_i$ from Excel.
4. Double-click *M*: set **Variance = 1**.
5. Set **Cov(M, any trait) = 0** (orthogonality).
6. Re-estimate; export standardized solution and residuals.

# Appendix D: Minimal LISREL Fragment

```
! Replace k, q, and w1..wk with your values
MO NY=k NE=q+1  LY=FU,FI  TE=SY  PS=SY  PH=SY
LA ETA1 ETA2 ... M
LY
 y1   1   0  ...  0
 y2   0   1  ...  0
```

```
...
 y1   0   0  ...  w1   ! Method column fixed
 y2   0   0  ...  w2
...
 yk   0   0  ...  wk
PH
 ETA1  *   0 ... 0
...
 M     0   0 ... 1
OU ND=3 RS MI
```

## Appendix E: Excel Workflow Checklist

- Paste standardized **residual correlation** matrix in the same item order.

- Enter item metadata (Reversed 0/1; Page; Order; Scale width; Polarity +1/−1; Length).

- Set and tune $\lambda$ until $R^2$ plateaus and weights stabilize.

- Use **Final_Weight** as fixed loadings for the method factor in AMOS/LISREL.

- Refit and compile the **effect stability panel**, **attribution table**, and **residual pockets** notes.